\documentclass[10pt]{iopart}
\usepackage{epsfig}
\begin{document}

\title[Longitudinal photocurrent spectroscopy of a single v-groove quantum wire]{Longitudinal photocurrent spectroscopy of a single GaAs/AlGaAs v-groove quantum wire}

\author{N I Cade\dag\footnote[2]{Present address: NTT Basic Research Laboratories, 3-1 Morinosato Wakamiya,
Kanagawa 243-0198, Japan. Email: ncade@will.brl.ntt.co.jp}, M Hadjipanayi\dag, R Roshan\dag, A C Maciel\dag, J F Ryan\dag, F Macherey\S, Th
Sch\"{a}pers\S  and H L\"{u}th\S}

\address{\dag\ Clarendon Laboratory, University of Oxford, Oxford, OX1 3PU, UK}

\address{\S\ Institut f\"{u}r Schichten und Grenzfl\"{a}chen, Forschungszentrum, 52425 J\"{u}lich, Germany}

\begin{abstract}
Modulation-doped GaAs v-groove quantum wires (QWRs) have been fabricated with novel electrical contacts made to two-dimensional
electron-gas (2DEG) reservoirs. Here, we present longitudinal photocurrent (photoconductivity / PC) spectroscopy measurements of a single
QWR. We  clearly observe conductance in the ground-state one-dimensional subbands; in addition, a highly temperature-dependent response is
seen from other structures within the v-groove. The latter phenomenon is attributed to the effects of structural topography and
localization on carrier relaxation. The results of power-dependent PC measurements suggest that the QWR behaves as a series of weakly
interacting localized states, at low temperatures.
\end{abstract}

\pacs{73.50.Pz, 73.63.Nm, 78.67.Lt}

\maketitle %\clearpage

\section{Introduction}
\label{PCintro}

In recent years there has been considerable effort expended in producing high quality quasi-one-dimensional (1D) electronic systems, in
order to study their fundamental physical properties. Various techniques have been developed in order to realize these systems in
semiconductors; these include the selective depletion of a high mobility two-dimensional electron-gas (2DEG) \cite{thornton86}, cleaved
edge overgrowth of quantum wells (QWs) \cite{pfeiffer90}, and vicinal growth on patterned substrates \cite{colas}. There is currently
increasing interest in the electronic transport properties of alternative quasi-1D materials, such as carbon nanotubes \cite{tans} and DNA
molecules \cite{fink}.

The optical properties of semiconductor v-groove quantum wire (QWR) structures have attracted extensive research, due to their potential
use in optoelectronic devices including quantum wire lasers \cite{sirigu00};  however, much less attention has been devoted to their
electronic properties. The possibility of constructing nanostructure device networks using low-dimensional interconnects necessitates a
greater understanding of carrier transport into and through such wires \cite{ferry}. Photocurrent (PC) spectroscopy provides a means of
directly investigating the excited state properties of a semiconductor structure. This is especially pertinent when injecting carriers from
nanoscale electrical contacts into low-dimensional structures. For the case of QWRs, PC spectroscopy provides information on the specific
real-space electronic transitions involved in transport, which cannot be obtained by standard conductance measurements.

Previous work in this area has mainly concentrated on \textit{transverse} (i.e.\ perpendicular to the wires) photocurrent measurements of
v-groove \mbox{p-i-n} structures \cite{Hamoudi,fu}, involving large ensembles of QWRs \cite{cingolani}.  These investigations provided
information on the energy levels within QWR structures but have given only limited insight into their overall electrical transport
properties. Saraydarov \emph{et al}.\ \cite{saraydarov} have recently reported longitudinal PC measurements of 24 wires that suffered from
significant parallel transport through the substrate and buffer layers.

In this paper we report temperature- and power-dependent \emph{longitudinal} photocurrent spectroscopy measurements of a \textit{single}
QWR. We have grown modulation-doped GaAs QWRs in  v-grooves with quasi- \{111\}A sidewalls. In order to electrically contact the wires,
wide grooves with an aperture of $\sim$50 $\mu $m have been fabricated at each end of the v-groove (see figure \ref{tem}(b)): as the wire
grows in the v-groove, a contiguous 2DEG is formed in the (100) bottom quantum well (BQW) which grows on the bottom facet of the wide
groove. This technique creates a tapering from 2D - 1D, and effectively eliminates the high contact resistance that occurs in sidewall
contact techniques due to a large wavefunction mismatch between the sidewall quantum well (SQW) and QWR \cite{lelarge}.

We have observed conductance in the ground-state one-dimensional subbands, and we have identified additional conductance features that
originate from other structures within the v-groove. Small changes in lattice temperature reveal a strong dependence of excited state
dynamics and relaxation on structural topography, and highlight the importance of carrier localization. We find that the power-dependence
of the PC response can be closely fit by a rate equation model for a two-level quantum dot system. This indicates that, at low
temperatures, the QWR behaves as a series of quasi-0D boxes rather than a quasi-1D system.

\section{Experimental details}\label{Experimental}

\begin{figure}[t!]
\begin{center}
\epsfig{file=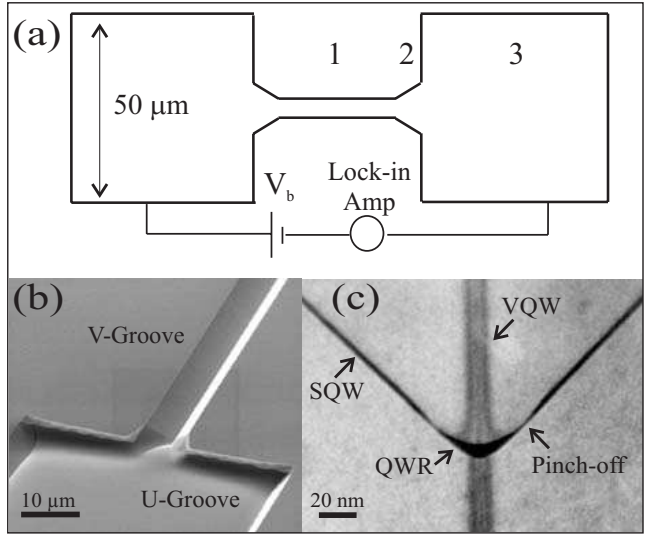,width=8.5cm}\end{center} \caption{(a) Schematic of a single groove (approximately to scale) showing the different
regions of interest : 1) v-groove ($\sim$25 $\mu$m length), 2) 1D-2D transition ($\sim$5 $\mu$m length), 3) end contact 2DEG. (b) SEM image
of one end of a v-groove showing the transition into the u-groove. (c) TEM cross-section of a v-groove showing the formation of a quantum
wire, side QWs and a Ga rich vertical QW.} \label{tem}
\end{figure}

A semi-insulating (100) GaAs substrate was patterned with an array of $\sim$25 $\mu$m long [01$\overline{1}$]-oriented v-grooves with 5
$\mu$m apertures; each groove was separated by 50 $\mu $m of Si0$_{2}$ mask. A nominal 2.5 nm GaAs quantum well (QW) was grown within
Al$_{0.3}$Ga$_{0.7}$As barriers by low-pressure metal-organic vapor-phase epitaxy. A 10 nm thickness of the barrier was modulation-doped 30
nm above (below) the QW with a Si concentration of 8 (4) $\times$ 10$^{17}$ cm$%
^{-3}$. Polycrystalline growth occurs between the grooves due to the residual Si0$_{2}$, which prevents the formation of a planar (100) QW
at the top of the SQWs and ensures that each groove is electrically isolated \cite{schapers}. Electrical contacts were made onto the 2DEG
regions that terminate each groove (figure \ref{tem}(b)).  Low temperature current-voltage (I-V) measurements indicate that the v-groove
structures are ohmic at low fields, and give a total wire resistance of $\sim$5 k$\Omega$. A detailed description of the growth and
characterization of these quantum wires is given elsewhere \cite{schapers}.

Figure \ref{tem}(a) shows a plan-view schematic of a single groove and the three different regions of interest: 1) the v-groove, 2) a 1D-2D
transition region, 3) a wide u-groove containing a 2DEG onto which electrical contacts are made. A cross-section of a v-groove is shown in
figure \ref{tem}(c), obtained by transmission electron microscopy (TEM); a crescent-shaped GaAs QWR forms at the vertex of two SQWs,
separated from them by a narrow constriction or pinch-off. A vertical quantum well (VQW) also forms at the center of the groove,
perpendicular to the substrate, due to the higher mobility of Ga atoms in the AlGaAs \cite{biasiol96}.

Photocurrent measurements were made using an Ar$^{+}$ pumped automatically-tuneable dye laser ($\sim $1.6 -- 1.8 eV), focussed to a spot
size of $\sim$35 $\mu$m on the v-groove surface. The sample was mounted on a variable temperature continuous flow cold-finger, with a base
temperature of 4.5 K. A DC voltage generator was used to apply a bias of a few hundred mV between the end contacts; by mechanically
chopping the laser-beam, the resultant photocurrent was obtained with standard lock-in techniques.

\section{Results and discussion}\label{Results}

\begin{figure}[t!]
\begin{center}
\epsfig{file=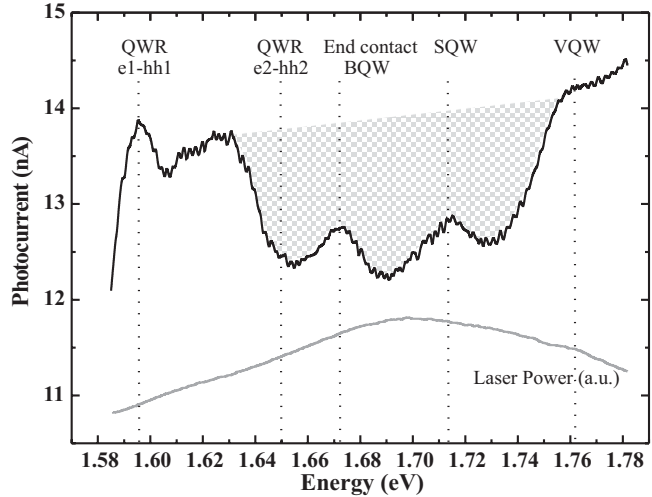,width=8.5cm}\end{center} \caption{PC spectrum (10 K) obtained with a maximum power of 150 $\protect\mu $W and a 80
mV bias. The shading indicates the region of reduced PC observed at low temperature. The dotted lines correspond to the positions of
luminescence features. The tuning curve for a typical scan of the dye laser is also shown.} \label{pc}
\end{figure}

In figure \ref{pc} we present a PC spectrum taken at 10 K. The dotted vertical lines indicate the positions of photoluminescence (PL)
features observed for the various groove structures  \cite{cade2}. The lines at 1.595 eV and 1.65 eV originate from e$_{1}$-hh$_{1}$ and
e$_{2}$-hh$_{2}$ (\textit{n}=1,2) recombination in the QWR respectively. These assignments have been confirmed by spatial mapping of the
luminescence from the v-groove, using near-field scanning optical microscopy \cite{marc,cade}, and also by theoretical modelling and
PL-excitation (PLE) measurements \cite{cade2}. A strong PC response can be seen at the $n=1$ QWR energy, indicating efficient exciton
generation and subsequent carrier transport in the ground state subbands. This is anomalous as the structure is supposedly heavily n-doped
and hence this transition should be forbidden; however, there is also no obvious absorption edge seen at the Fermi energy in PLE
measurements taken for this sample. The issue of modulation doping in these v-groove structures is complicated and subject to ongoing
research. It is possible that the transfer of free carriers into the QWR is inefficient under certain growth or experimental conditions.

There is a prominent drop in the overall PC response between the low energy side of the \textit{n}=2 QWR transition and the VQW energy, as
indicated by the shaded region in figure \ref{pc}. This is evidence of reduced extraction efficiency of electron-hole pairs through the
device at low temperature, as discussed below. Absorption is also clearly evident at the SQW energy, which is consistent with parallel
conduction mechanisms; at low temperatures carrier relaxation from the sidewalls into the QWR is greatly restricted by the pinch-off
regions \cite{cade2}. Despite the large area presented by the SQWs for conduction, the PC response at this energy is relatively small. This
is due to discontinuity of the \{111\}A facet at the ends of the groove, which prevents efficient carrier transfer between the end BQWs and
the SQWs (see figure \ref{tem}(b)). Another strong peak occurs at the VQW energy; this is indicative of carrier transfer between the VQW
and QWR \cite{kiener}, as parallel conduction is unlikely due to strong compositional disorder in the VQW.

\subsection{Temperature characteristics}\label{pctempsec}

\begin{figure}[t!]
\begin{center}
\epsfig{file=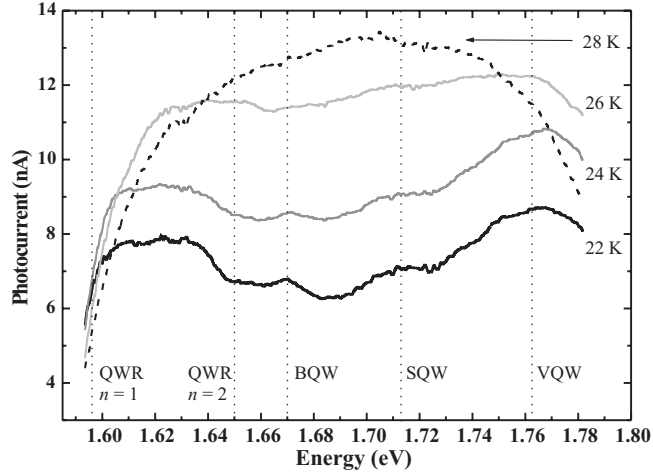,width=8.5cm}\end{center} \caption{PC spectra for a single groove at different temperatures (200 mV, 30 $\mu$m spot,
maximum power 4 $\mu$W). The dotted lines correspond to the features in figure \ref{pc}.} \label{pctemp}
\end{figure}

PC spectra were recorded at different temperatures and are presented in figure \ref{pctemp}. With a small rise in
temperature the absorption features disappear, and the PC increases greatly in the middle energy range. This is
especially noticeable at the \textit{n}=2 QWR energy.

In the QWR, the e$_{2}$ electronic wavefunction is localized over the \{311\}A facets which create the pinch-off \cite{wang1}. These facets
are subject to large interfacial disorder, which causes strong exciton localization \cite{bellessa}, and greatly inhibited relaxation
\cite{oberli}. These issues are  discussed in detail elsewhere \cite{cade2}. At very low temperatures, localization effects play a dominant
role in limiting carrier transport; this results in a reduction of the photocurrent contribution from certain parts of the structure, as
seen in figure \ref{pc} between 1.63 eV and 1.75 eV. The radiative recombination rate for an exciton is found to be proportional to its
coherence volume in the confining potential \cite{feldmann}. Excitons localized over lengths $L$, comparable to the excitonic Bohr radius
($a_{\rm B}$) have a relatively long lifetime of up to 500 ps at low temperatures \cite{bellessa}, and are thus susceptible to dissociation
by hot carriers. In the limits of $L$$\gg$$a_{\rm B}$ and $L$$\ll$$a_{\rm B}$, the radiative lifetime tends to that of a free exciton,
$\sim$150 ps \cite{citrin}.  The \textit{n}=1 and \textit{n}=2 QWR states are expected to have different localization lengths as their
respective wavefunctions are centered on crystal facets with different interfacial roughness \cite{wang1}. Thus, at low temperature
excitons in the \textit{n}=2 QWR state may recombine before they are dissociated, and are not able to contribute to the photocurrent. As
the lattice temperature is increased, these excitons are de-trapped and can relax more efficiently to the QWR ground-state where they are
ionized within the recombination time. This results in an increase in the PC at the \textit{n}=2 QWR energy, as seen in figure
\ref{pctemp}. The PC at the SQW energy also increases with temperature due to a real-space transfer of carriers into the QWR
\cite{cade2,cade}, which results in increased carrier-carrier scattering and smearing of the absorption features.

In this discussion we have assumed that radiative recombination is the dominant mechanism, as non-radiative processes are expected to
become significant only at temperatures greater than the range considered here \cite{akiyama}. We have performed time-resolved measurements
on similar QWR structures and obtain low temperature lifetimes of $\sim $350 ps and $\sim $150 ps for the \textit{n}=1 and \textit{n}=2
states respectively \cite{cade2}, which supports the above conclusions. Above 30 K, we observe a decrease in the relaxation time of the
\textit{n}=2 state to the ground state, which is also consistent with these measurements.

\subsection{Power characteristics}\label{pcpowersec}

\begin{figure}[t!]
\begin{center}
\epsfig{file=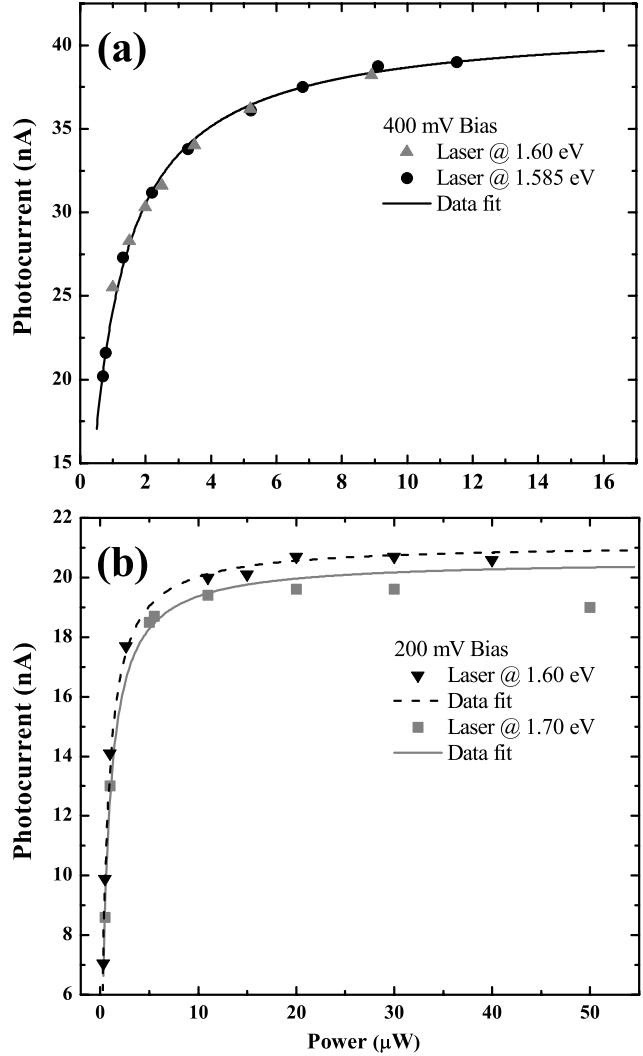,width=8.5cm}\end{center} \caption{Photocurrent as a function of laser power, for different laser energies at 10 K;
(a) 400 mV bias across the groove, (b) 200 mV bias, for higher laser powers. The data have been fitted with a function based on a rate
equation model described in the text.} \label{pcpower}
\end{figure}

In figure \ref{pcpower} we show the PC response as a function of laser power, when illuminating at various
energies. In all cases the PC increases rapidly with a small increase in power, and then saturates at a value that
depends on the bias.

The high sensitivity of the PC to low powers explains why the spectrum in figure \ref{pc} shows a clear peak at $\sim$ 1.6 eV
while those in figure \ref{pctemp} do not. The laser is at the extreme of its tuning range at this energy, and the excitation power (and
hence PC response) can vary significantly on different days depending on the fine tuning of the laser.

Beham \textit{et al}.\ \cite{beham02} have observed a similar power-dependent PC response in single InGaAs quantum dots (QDs), which they
attributed to bleaching of the dot ground-state. We outline their model below and discuss its relevance to a 1D wire system.

The saturation of the PC in a QD with increasing excitation power can be described by a fundamental rate equation model for a two-level
system. This assumes the dot is in level 1 if it is unoccupied, and in level 2 if it is occupied with one exciton. Stimulated processes are
assumed to occur from level 2 with a probability $M$, and radiative recombination will occur in a time $\tau_{\rm r}$. Excitons can also
escape out of the dot by field-dependent carrier tunnelling, with a escape time $\tau_{\rm esc}(F)$. This leads to occupation-dependent
absorption behavior in the QD, which produces a nonlinear power dependence of the PC resulting in saturation. At higher electric fields the
tunnelling time is reduced and the QD returns faster to its initial empty state; therefore, a higher absorption rate is possible leading to
a higher saturated PC. With an excitation power $P$, the steady-state photocurrent is found to be
\begin{equation}\label{PCeq}
  I_{PC}=\left(\frac{e}{2\tau_{\rm esc}}\right)\frac{P}{P+(1/2M)(1/\tau_{\rm esc}+1/\tau_{\rm r})}.
\end{equation}

This expression has been used to fit the data points in figure \ref{pcpower}; we assume $\tau_{\rm r}\equiv\tau_{\rm loc}\simeq350$ ps
\cite{cade2}. From the data in figure \ref{pcpower}(a), with resonant excitation of the QWR at 1.585 eV, we obtain values of $M=0.36$ and
an escape time $\tau_{\rm esc}=1.9$ ps for 400 mV bias (80 V\,cm$^{-1}$). At 200 mV bias, the data fits in figure \ref{pcpower}(b) give
$M=0.24$ (0.20) and $\tau_{\rm esc}=3.8$ (3.9) ps for illumination at 1.6 eV (1.7 eV).

The closeness of the fits, at least over low powers, suggests that the model developed for QDs is still valid in our 1D QWR system. As
discussed above, this is likely to be a result of localization effects at low temperatures; monolayer growth fluctuation will generate a
series of quasi-0D boxes along the wire axis \cite{hasen97}, thus emulating the isolated QD model. The escape time at 400 mV bias is half
the value at 200 mV, which suggests that in our case $\tau_{\rm esc}^{-1}$ gives an indication of the rate of dissociation of localized
excitons by field-accelerated carriers. These values of $\tau_{\rm esc}$ are much smaller than those obtained by Beham \textit{et al}.\
even though our fields are 1000 times less; this is likely to be due to the relatively high density of '0D' localization sites along the
wire ($\sim5 \times 10^{4}$ cm$^{-1}$ \cite{cade2}) and the presence of a 1D delocalized exciton continuum.

Figure \ref{pcpower}(b) shows that the saturation PC is significantly smaller when exciting at 1.70 eV rather than quasi-resonantly with
the $n=1$ state at 1.6 eV. This is consistent with the discussion above; at low temperatures, excitons generated in excited states of the
wire recombine more rapidly than ground state excitons, thus generating fewer dissociated free carriers.

There is not sufficient data to comment on the field- and excitation energy-dependence of $M$, and a detailed analysis of the true meaning
of $M$ and $\tau_{\rm esc}$ in a 1D system will require a more comprehensive investigation. The deviation of the PC from a saturation
value, observed at higher powers in figure \ref{pcpower}(b), may be due to the presence of the 1D continuum. Scattering with phonons and
other carriers may be significant at higher powers and has also been neglected in the model.

\section{Conclusion}\label{Conc}
In summary, we have presented longitudinal photocurrent measurements of a single v-groove quantum wire. The
photocurrent response from certain structures within the v-groove is found to be highly temperature dependent. In
particular, a marked drop is observed in the contribution from the first excited state of the quantum wire, below
30 K. This has been attributed to frustrated carrier relaxation resulting from localization effects.

The total photocurrent through the v-groove has been measured as a function of incident laser power; the PC
response can be closely fit by a rate equation model for a two-level quantum dot system. This is consistent with
the scenario that the QWR behaves as a series of quasi-0D boxes, at low temperatures.

This work was supported by the EPSRC (UK), and the EC through the ULTRAFAST network.

\end{document}